\documentclass[12pt]{article}
\usepackage[cp1251]{inputenc}
\usepackage[english,russian]{babel}
\usepackage{indentfirst}
\usepackage{graphicx}
\usepackage{amsmath}
\usepackage{amssymb}
\input{epsf}

\newcommand{\eps}{\varepsilon}
\newcommand{\vv}{\mathbf}
\newcommand{\dd}{\partial}

\title{Ion-Acoustic Solitons in Bi-Ion Dusty Plasma}
\author{V. V. Prudskikh}
\date{}

\begin{document}
\maketitle
\noindent

The propagation of ion-acoustic solitons in a warm dusty plasma containing two ion species is investigated theoretically. Using an approach based on the Korteveg-de-Vries equation, it is shown that the critical value of the negative ion density that separates the domains of existence of compressi-\\on and rarefaction solitons depends continuously on the dust density. A modified Korteveg-de Vries equation for the critical density is derived in the higher order of the expansion in the small parameter. It is found that the nonlinear coefficient of this equation is positive for any values of the dust density and the masses of positive and negative ions. For the case where the negative ion density is close to its critical value, a soliton solution is found that takes into account both the quadratic and cubic nonlinearities. The propagation of a solitary wave of arbitrary amplitude is investigated by the quasi-potential method. It is shown that the range of the dust densities around the critical value within which solitary waves with positive and negative potentials can exist simultaneously is relatively wide.

\bigskip\noindent {\tt PACS: 52.35.--g, 52.35.--Fp}

\section{INTRODUCTION}

It is well known that the presence of negative ions in plasma substantially modifies ion-acoustic solitons described by the Korteveg-de-Vries (KdV) equation. This is related to the fact that the nonlinear coefficient of the equation can vanish at a certain critical value of the negative ions density. If this density is below the critical value, then the solution has the form of a solitary compression wave; otherwise, it has the form of a rarefaction wave. To describe the nonlinear properties of an ion-acoustic wave at negative ion densities close to the critical density, it is necessary to take into account the higher orders of the expansion in the small parameter. In doing so, Watanabe [1] arrived at a modified KdV (MKdV) equation, which was then analyzed in [2]--[5]. Mishra and Chhabra [6] showed that, in a plasma with a finite ion temperature, there can exist two types of ion-acoustic waves---fast and slow ones. In [7], the influence of nonisothermal electrons on the properties of ion-acoustic solitons in plasmas containing positive and negative ions was studied by the quasi-potential method. The propagation of nonlinear compression and rarefaction ion-acoustic waves in a bi-ion plasma was studied by McKenzy et al. [8]--[9] in a gas-dynamic approach, which is alternative to the quasi-potential method.

In the present work, the propagation of ion-acoustic solitons in a dusty plasma containing both positive and negative ions is investigated theoretically. It is well known [10] that, when a plasma with one positive ion species contains a negatively charged dust, there is a critical dust density below (above) which an ion-acoustic soliton has the form of a compression (rarefaction) wave. When the plasma also contains negative ions, the problem has two independent parameters--- the negative ion density and the density of dust, which, depending on the external conditions, can be charged ether negatively or positively. As a result, at a given dust density, the critical value of the negative ion density can be shifted or can even be absent at all. Moreover, for given masses and temperatures of the plasma ion components, the critical value of the negative ion density depends continuously on the dust density. Therefore, the nonlinear coefficient in the KdV equation vanishes on a continuous line in the "`dust density--negative ion density"' plane. This line separates the domains of existence of compression and rarefaction ion-acoustic solitons.

The vanishing of the nonlinear coefficient in the KdV equation at the points corresponding to the critical density unambiguously relates the dust density to the negative ion density. In a dusty plasma, the nonlinear coefficient in the MKdV equation is a function of one of the above parameters (e.g., of the dust density) rather than merely a number. Therefore, it is interesting to examine whether this coefficient can actually vanish and change its sign. The vanishing of the nonlinear coefficient in the MKdV equation would mean the existence of a double critical point at which of both the first- and second-order nonlinearities disappear. To find out whether the nonlinear coefficient changes its sign is important for understanding the properties of solutions to the MKdV equation. When the coefficient is positive, the solution has the form of a soliton. In contrast, when it is negative, a double layer can form in the system. The results of our study show that the nonlinear coefficient is positive at any values of the dust density and the masses of positive and negative ions. Therefore, in a bi-ion dusty plasma, the solution near the critical points has the form of a soliton.

In studying the propagation of an ion-acoustic soliton in a dusty plasma, we will assume that the charge of a dust grain in the soliton field remains constant. This is true if the charging frequency $\nu_{ch}=a\omega_{pi}/\sqrt{2\pi}\lambda_{Di}$ ($a$ -- is the dust grain radius, $\omega_{pi}=(4\pi e^2n_{i0}/m_i)^{1/2}$ -- is the plasma ion frequency, and $\lambda_{Di}=(T_i/4\pi e^2n_{i0})^{1/2}$ -- is the ion Debye radius) is considerably smaller than the reciprocal of the time during which the soliton passes through a given point in the plasma. This time can be estimated as $\tau_i\sim L/c_s\sim N\lambda_{Di}/c_s=N\omega_{pi}^{-1}T_i/T_e\leq N\omega_{pi}^{-1}$, where $L$ -- is the soliton length and $N=L/\lambda_{Di}$. The condition for the grain charge to be constant is then satisfied when $N\nu_{ch}/\omega_{pi}\sim Na/\lambda_{Di}\ll 1$. For $a\sim 10^{-4}$ cm, $T_i\sim 1 $ eV, and $N=100$, the above assumption is valid if $10^{-5}n_{i0}^{1/2}\ll 1$, which corresponds to a fairly low plasma density ($n_{i0}<10^6$).  In a dense plasma, the charge of dust grains in the field of an ion-acoustic soliton cannot be considered constant. In this case, either ion-acoustic shocks [11] or weakly-dissipative ion-acoustic dust solitons [12] can form, depending on the electron distribution function.

The present paper is organized as follows. In Section 2, we write out a set of basic dimensionless equations. In Section 3, the KdV equation is derived using the method of perturbations and the dependence of the amplitude of a solitary wave on the plasma parameters is investigated numerically. In Section 4, the MKdV equation is derived, its nonlinear coefficient is analyzed, and a solution is obtained that describes an ion-acoustic wave in a plasma at densities close the critical density. In Section 5, the propagation of a wave of arbitrary amplitude is investigated by the quasi-potential method. Finally, in Section 6, the main results of this study are summarized.

\section{BASIC EQUATIONS}

Let us consider a dusty plasma containing positive and negative ions. For definiteness, we assume that the ions are singly charged. The problem to be solved is described by the set of equations that includes the equations of continuity and motion for both ion components, as well as Poisson's equation:\begin{equation}
\frac{\dd n_1}{\dd t}+\frac{\dd}{\dd x}\left(n_1v_1\right)=0,
\end{equation}
\begin{equation}
\left(\frac{\dd}{\dd t}+v_1\frac{\dd}{\dd x}\right)v_1=-\frac{\dd\phi}{\dd x}-3\theta_1n_1\frac{\dd n_1}{\dd x},
\end{equation}
\begin{equation}
\frac{\dd n_2}{\dd t}+\frac{\dd}{\dd x}\left(n_2v_2\right)=0,
\end{equation}
\begin{equation}
\left(\frac{\dd}{\dd t}+v_2\frac{\dd}{\dd x}\right)v_2=\frac{1}{Q}\left(\frac{\dd\phi}{\dd x}-3\theta_2n_2\frac{\dd n_2}{\dd x}\right),
\end{equation}
\begin{equation}
\frac{\dd^2\phi}{\dd x^2}=e^{\phi}-\mu_1n_1+\mu_2n_2+\mu.
\end{equation}
Here, the coordinate $x$,  the time $t$, and the velocities $v_1, v_2$ are normalized to the electron Debye radius $\lambda_D=(4\pi n_{e0}e^2/T_e)^{1/2}$, the inverse plasma frequency $\omega_{p}^{-1}=(m_1/4\pi n_{e0}e^2)^{1/2}$, and the ion-acoustic velocity $c_s=(T_e/m_1)^{1/2}$, respectively; $\mu_1=n_{i10}/n_{e0}$, $\mu_2=n_{i20}/n_{e0}$, $\phi=e\phi'/T_e$, $\theta_1=T_{i1}/T_e$, $\theta_2=T_{i2}/T_e$, $Q=m_2/m_1$, $T_e, T_{i1}, T_{i2}$ are the temperatures of electrons and positive and negative ions, respectively. The densities of electrons and positive and negative ions are normalized to their equilibrium values, $n_{e0}, n_{i10}, n_{i20}$, respectively. We assume that the wave propagates adiabatically, so the electrons density is $n_e=\exp(\phi)$. The parameters $\mu_1$ and $\mu_2$ are related by the quasineutrality condition $\mu_1=1+\mu_2+\mu$, where $\mu=Zn_{d0}/n_{e0}$, $Z$ is the charge number of a dust grain, and $n_{d0}$ is the dust number density.

\section{KORTEVEG-DE-VRIES EQUATION}

We will search for the solution to the above set of equations in the form
\begin{equation}
\vv{A}=\vv{A}_0+\eps\vv{A}_1+\eps^2\vv{A}_2+\eps^3\vv{A}_3+...,
\end{equation}
where $\vv{A}=(n_1,v_1,n_2,v_2,\phi)$, $\vv{A}_0=(1,0,1,0,0)$. To derive the Korteveg-de-Vries equation, we make use of the standard method of coordinate stretching,
\begin{equation}
\xi=\eps^{1/2}(x-Vt),\,\,\,\,\tau=\eps^{3/2}t.
\end{equation}

In the first order in $\eps$, the set of equations is written as
\begin{equation}
-V\frac{\dd n_{11}}{\dd\xi}+\frac{\dd v_{11}}{\dd\xi}=0,
\end{equation}
\begin{equation}
-V\frac{\dd v_{11}}{\dd\xi}+\frac{\dd\phi_1}{\dd\xi}+3\theta_1\frac{\dd n_{11}}{\dd\xi}=0,
\end{equation}
\begin{equation}
-V\frac{\dd n_{21}}{\dd\xi}+\frac{\dd v_{21}}{\dd\xi}=0,
\end{equation}
\begin{equation}
-V\frac{\dd v_{21}}{\dd\xi}-\frac{1}{Q}\frac{\dd\phi_1}{\dd\xi}+3\frac{\theta_2}{Q}\frac{\dd n_{21}}{\dd\xi}=0.
\end{equation}
\begin{equation}
\phi_1-\mu_1n_{11}+\mu_2n_{21}=0.
\end{equation}
Integrating Eqs. (8)-(11) with the initial conditions $v_{11}(\xi\rightarrow\pm\infty)=0,v_{21}(\xi\rightarrow\pm\infty)=0$, $n_{11}(\xi\rightarrow\pm\infty)=0, n_{21}(\xi\rightarrow\pm\infty)=0$ and substituting the expressions for $n_{11}$ and $n_{21}$ into Eq. (12), we obtain the following expressions for the first-order quantities, as well as the relationship that determines the linear velocity of the wave:
\begin{gather}
n_{11}=\frac{\phi_1}{V^2-3\theta_1},\hspace{10 mm} v_{11}=\frac{V}{V^2-3\theta_1}\phi_1,\nonumber\\
n_{21}=-\frac{\phi_1}{QV^2-3\theta_2},\hspace{10 mm} v_{21}=-\frac{V}{QV^2-3\theta_2}\phi_1,\\
\frac{\mu_1}{V^2-3\theta_1}+\frac{\mu_2}{QV^2-3\theta_2}=1.\nonumber
\end{gather}

In the second order in $\eps$, we have
\begin{equation}
\frac{\dd n_{11}}{\dd\tau}-V\frac{\dd n_{12}}{\dd\xi}+\frac{\dd v_{12}}{\dd\xi}+\frac{\dd}{\dd\xi}(n_{11}v_{11})=0,
\end{equation}
\begin{equation}
\frac{\dd v_{11}}{\dd\tau}-V\frac{\dd v_{12}}{\dd\xi}+v_{11}\frac{\dd v_{11}}{\dd\xi}+\frac{\dd\phi_2}{\dd\xi}+3\theta_1\frac{\dd n_{12}}{\dd\xi}+3\theta_1n_{11}\frac{\dd n_{11}}{\dd\xi}=0,
\end{equation}
\begin{equation}
\frac{\dd n_{21}}{\dd\tau}-V\frac{\dd n_{22}}{\dd\xi}+\frac{\dd v_{22}}{\dd\xi}+\frac{\dd}{\dd\xi}(n_{21}v_{21})=0,
\end{equation}
\begin{equation}
\frac{\dd v_{21}}{\dd\tau}-V\frac{\dd v_{22}}{\dd\xi}+v_{21}\frac{\dd v_{21}}{\dd\xi}-\frac{1}{Q}\frac{\dd\phi_2}{\dd\xi}+\frac{3\theta_2}{Q}\frac{\dd n_{22}}{\dd\xi}+\frac{3\theta_2}{Q}n_{21}\frac{\dd n_{21}}{\dd\xi}=0,
\end{equation}
\begin{equation}
\frac{\dd^2\phi_1}{\dd\xi^2}=\phi_2+\phi_1^2/2-\mu_1n_{12}+\mu_2n_{22}.
\end{equation}
Excluding the term $\dd v_{12}/\dd\xi$ from Eqs. (14) and (15) and using Eqs. (13), we obtain
\begin{equation}
\frac{\dd n_{12}}{\dd\xi}=\frac{1}{V^2-3\theta_1}\frac{\dd\phi_2}{\dd\xi}+\frac{2V}{(V^2-3\theta_1)^2}\frac{\dd\phi_1}{\dd\tau}+\frac{3(V^2+\theta_1)}{(V^2-3\theta_1)^3}\phi_1\frac{\dd\phi_1}{\dd\xi}.
\end{equation}
Analogously, we derive the expression for $\dd n_{22}/\dd\xi$:
\begin{equation}
\frac{\dd n_{22}}{\dd\xi}=-\frac{1}{QV^2-3\theta_2}\frac{\dd\phi_2}{\dd\xi}-\frac{2QV}{(QV^2-3\theta_2)^2}\frac{\dd\phi_1}{\dd\tau}+\frac{3(QV^2+\theta_2)}{(QV^2-3\theta_2)^3}\phi_1\frac{\dd\phi_1}{\dd\xi}.
\end{equation}
Differentiating Eq. (18) over $\xi$ and using expressions (19) and (20), we obtain the KdV equation,
\begin{equation}
\frac{\dd\phi_1}{\dd\tau}+A\phi_1\frac{\dd\phi_1}{\dd\xi}+\frac{b}{2}\frac{\dd^3\phi_1}{\dd\xi^3}=0,
\end{equation}
where
\begin{displaymath}
A/b=\frac{1}{2}\left[\frac{3\mu_1(V^2+\theta_1)}{(V^2-3\theta_1)^3}-\frac{3\mu_2(QV^2+\theta_2)}{(QV^2-3\theta_2)^3}-1\right],\hspace{5 mm} b=\frac{1}{V}\left[\frac{\mu_1}{(V^2-3\theta_1)^2}+\frac{Q\mu_2}{(QV^2-3\theta_2)^2}\right]^{-1}.
\end{displaymath}

After switching to the coordinate $\eta=\xi-u\tau$, the solution to Eq. (21) that satisfies the condition $\phi_1(\xi\rightarrow\pm\infty)=0$ can be rewritten as
\begin{equation}
\phi_1=\phi_m\ch^{-2}((u/2b)^{1/2}\eta),\,\,\,\phi_m=3u/A.
\end{equation}

For numerical analysis, it is convenient to introduce the dimensionless coefficients  $\alpha=\mu_2/\mu_1=n_{i20}/n_{i10}$ and $\beta=Zn_{d0}/n_{i10}$, which are the density ratio between the negative and positive ions and the ratio of the dust grain charge to the charge of a positive ion, respectively. Figure 1 shows how the wave amplitude depends on $\alpha$ at different values of $\beta$. It is seen that the critical density of negative ions decreases in the presence of a negatively charged dust. In contrast, if the dust is charged positively, then the critical density increases. If the dust is charged negatively, then, for a given value of the parameter $\alpha$, the soliton amplitude in a dusty plasma is higher than that in the absence of dust, and vice versa, if the dust is charged positively, then the soliton amplitude in a dusty plasma is lower than that in a bi-ion plasma without dust.

\section{MODIFIED KORTEVEG-DE-VRIES EQUATION}

Let us now assume that the density ratio between the negative and positive ions is equal to its critical value, i.e., $A=0$. In this case, in order to describe the nonlinear properties of the medium, it is necessary to take into account the higher orders of the expansion in the small parameter. We again apply the method of coordinate stretching,
\begin{equation}
\xi=\eps(x-Vt),\hspace{2 cm}\tau=\eps^3t.
\end{equation}
In the first order in $\eps$, the equations remain unchanged. In the second order in $\eps$, the $\dd/\dd\tau$ derivative in Eqs. (14)--(17) is absent. This allows us to integrate relationships (19) and (20) analytically. Using continuity equations (14) and (16), we also obtain expressions for $v_{21}$ и $v_{22}$ that are necessary to construct the theory in the third order in $\eps$:
\begin{gather}
n_{12}=\frac{\phi_2}{V^2-3\theta_1}+\frac{3(V^2+\theta_1)}{2(V^2-3\theta_1)^3}\phi_1^2,\hspace{10 mm} v_{12}=\frac{V}{V^2-3\theta_1}\phi_2,+\frac{V(V^2+9\theta_1)}{2(V^2-3\theta_1)^3}\phi_1^2,\nonumber\\
n_{22}=-\frac{\phi_2}{QV^2-3\theta_2}+\frac{3(QV^2+\theta_2)}{2(QV^2-3\theta_2)^3}\phi_1^2,\hspace{10 mm} v_{22}=-\frac{V}{QV^2-3\theta_2}\phi_2,+\frac{V(QV^2+9\theta_2)}{2(QV^2-3\theta_2)^3}\phi_1^2.
\end{gather}

In the third order in $\eps$, Eqs. (1)--(4) take the form
\begin{equation}
\frac{\dd n_{11}}{\dd\tau}-V\frac{\dd n_{13}}{\dd\xi}+\frac{\dd v_{13}}{\dd\xi}+\frac{\dd}{\dd\xi}(n_{12}v_{11}+n_{11}v_{12})=0,
\end{equation}
\begin{equation}
\frac{\dd v_{11}}{\dd\tau}-V\frac{\dd v_{13}}{\dd\xi}+\frac{\dd}{\dd\xi}(v_{11}v_{12})+\frac{\dd\phi_3}{\dd\xi}+3\theta_1\frac{n_{13}}{\dd\xi}+3\theta_1\frac{\dd}{\dd\xi}(n_{11}n_{12})=0,
\end{equation}
\begin{equation}
\frac{\dd n_{21}}{\dd\tau}-V\frac{\dd n_{23}}{\dd\xi}+\frac{\dd v_{23}}{\dd\xi}+\frac{\dd}{\dd\xi}(n_{22}v_{21}+n_{21}v_{22})=0,
\end{equation}
\begin{equation}
\frac{\dd v_{21}}{\dd\tau}-V\frac{\dd v_{23}}{\dd\xi}+\frac{\dd}{\dd\xi}(v_{21}v_{22})-\frac{1}{Q}\frac{\dd\phi_3}{\dd\xi}+\frac{3\theta_2}{Q}\frac{n_{23}}{\dd\xi}+\frac{3\theta_2}{Q}\frac{\dd}{\dd\xi}(n_{21}n_{22})=0.
\end{equation}
Excluding the terms with $\dd v_{13}/\dd\xi$ и $\dd v_{23}/\dd\xi$, we obtain
\begin{gather}
\frac{\dd n_{13}}{\dd\xi}=\frac{1}{V^2-3\theta_1}\frac{\dd\phi_3}{\dd\xi}+\frac{2V}{(V^2-3\theta_1)^2}\frac{\dd\phi_1}{\dd\tau}+\frac{3(V^2+\theta_1)}{(V^2-3\theta_1)^3}\frac{\dd}{\dd\xi}(\phi_1\phi_2)+\nonumber\\
+\frac{3(5V^4+30V^2\theta_1+9\theta_1^2)}{2(V^2-3\theta_1)^5}\phi_1^2\frac{\dd\phi_1}{\dd\xi}.
\end{gather}
\begin{gather}
\frac{\dd n_{23}}{\dd\xi}=-\frac{1}{QV^2-3\theta_2}\frac{\dd\phi_3}{\dd\xi}-\frac{2QV}{(QV^2-3\theta_2)^2}\frac{\dd\phi_1}{\dd\tau}+\frac{3(QV^2+\theta_2)}{(QV^2-3\theta_2)^3}\frac{\dd}{\dd\xi}(\phi_1\phi_2)-\nonumber\\
-\frac{3(5Q^2V^4+30QV^2\theta_2+9\theta_2^2)}{2(QV^2-3\theta_2)^5}\phi_1^2\frac{\dd\phi_1}{\dd\xi}.
\end{gather}
Differentiating Poisson's equation over $\xi$ in the third order in $\eps$ yields
\begin{equation}
\frac{\dd^2\phi_1}{\dd\xi^2}=\phi_3+\phi_1\phi_2+\phi_1^3/6-\mu_1n_{13}+\mu_2n_{23}.
\end{equation}
Then, substituting expressions (29) and (30) into Eq. (31), we arrive at MKdV equation
\begin{equation}
\frac{\dd\phi_1}{\dd\tau}+B\phi_1^2\frac{\dd\phi_1}{\dd\xi}+\frac{b}{2}\frac{\dd^3\phi_1}{\dd\xi^3}=0,
\end{equation}
where
\begin{displaymath}
B/b=\frac{1}{4}\left[\frac{3\mu_1(5V^4+30V^2\theta_1+9\theta_1^2)}{(V^2-3\theta_1)^5}+\frac{3\mu_2(5Q^2V^4+30QV^2\theta_2+9\theta_2^2)}{(QV^2-3\theta_2)^5}-1\right].
\end{displaymath}
Switching to the coordinate $\eta=\xi-u\tau$ and integrating Eq. (32) under the condition  $\phi_1(\xi\rightarrow\pm\infty)=0$, we obtain a solution in the form of a solitary wave,
\begin{equation}
\phi_1=\pm\phi_m\ch^{-1}((2u/b)^{1/2}\eta),\,\,\,\phi_m=(6u/B)^{1/2}.
\end{equation}

Let us now assume that the negative ion density is close to its critical value but is not exactly equal to it. Then, the second-order terms in Poisson's equation yield
\begin{gather}
n_{e2}-\mu_1n_{12}+\mu_2n_{22}=\phi_2+\phi_1^2/2-\mu_1\left[\frac{\phi_2}{V^2-3\theta_1}+\frac{3(V^2+\theta_1)}{2(V^2-3\theta_1)^3}\phi_1^2\right]+\nonumber\\
+\mu_2\left[\frac{\phi_2}{QV^2-3\theta_2}+\frac{3(QV^2+\theta_2)}{2(QV^2-3\theta_2)^3}\phi_1^2\right]=-A\phi_1^2.
\end{gather}
If $A=O(\eps)$,  then we have $A\phi_1^2=O(\eps^3)$. Accordingly, there is no terms of the second-order in $\eps$ on both the left-hand and right-hand sides of Poisson's equation, because, although each of the terms $\phi_2, \phi_1^2, n_{12}$ and $n_{22}$ is on the order of $\eps^2$, their sum is a quantity of the third order in $\eps$. In this case, instead of Eq. (32), we obtain the equation
\begin{equation}
\frac{\dd\phi_1}{\dd\tau}+A\phi_1\frac{\dd\phi_1}{\dd\xi}+B\phi_1^2\frac{\dd\phi_1}{\dd\xi}+\frac{b}{2}\frac{\dd^3\phi_1}{\dd\xi^3}=0,
\end{equation}
where the second and the third terms on the left-hand side are of the same order (note that, in deriving this equation, the term $A\cdot\dd(\phi_1\phi_2)/\dd\xi$ was omitted because it is of the higher order than the other terms).

In the terms of the coordinate $\eta=\xi-u\tau$ , the solution to Eq. (35) that has the form of a solitary wave and satisfies the condition $\phi_1(\xi\rightarrow\pm\infty)=0$ is expressed as
\begin{equation}
\phi_1=\frac{6u/A}{1\pm\sqrt{1+6Bu/A^2}ch[(2u/b)^{1/2}\eta]}.
\end{equation}
The amplitude of the soliton is
\begin{equation}
\phi_m=\frac{6u/A}{1\pm\sqrt{1+6Bu/A^2}}.
\end{equation}
If  $B=0$ or $A=0$ , then expressions (36) and (37) transform into formula (22) or (33), respectively.

As was mentioned above, the critical value of the negative ions density is unambiguously related to the dust density. The value of $\beta$ at which the nonlinear coefficient $A$ in Eq. (21) is zero is plotted in Fig. 2 as a function of the parameter $\alpha$ for different mass ratios between negative and positive ions. The points at which the vertical line $\beta$  intersects the plots correspond to the critical densities of negative ions in a plasma without dust. Figure 3 shows the amplitude of an MKdV soliton as a function of  $\beta$  at different values of $Q$. A numerical analysis shows that the coefficient $B$ is positive at any value of $Q$. Therefore, in the vicinity of $A=0$, Eq. (35) describes solitary waves and has no solutions in the form of a double layer.

\section{ANALYSIS OF THE QUASI-POTENTIAL}

Let us consider the propagation of a wave of arbitrary amplitude. Passing to the frame of reference related to the wave, $\eta=x-Mt$ (where $M$ is the Mach number in terms of $c_s$); integrating Eqs. (1)--(4) and substituting the resulting expressions for the ions densities into Poisson's equation, we obtain
\begin{gather}
\frac{1}{2}\left(\frac{d\phi}{d\eta}\right)^2+U(\phi)=0,\\
U(\phi)=1-e^{\phi}+\frac{\mu_1\sigma^3_{1+}}{6\sqrt{3\theta_1}}\left[1-\left(1-\frac{2\phi}{\sigma^2_{1+}}\right)^{3/2}-\frac{\sigma^3_{1-}}{\sigma^3_{1+}}\left(1-\frac{2\phi}{\sigma^2_{1-}}\right)^{3/2}\right]+\nonumber\\
+\frac{\mu_2\sigma^3_{2+}}{6\sqrt{3\theta_2}}\left[1-\left(1+\frac{2\phi}{\sigma^2_{2+}}\right)^{3/2}-\frac{\sigma^3_{2-}}{\sigma^3_{2+}}\left(1+\frac{2\phi}{\sigma^2_{2-}}\right)^{3/2}\right]-\mu\phi,\nonumber
\end{gather}
where $\sigma_{1\pm}=M\pm\sqrt{3\theta_1}$, $\sigma_{2\pm}=\sqrt{Q}M\pm\sqrt{3\theta_2}$.

Expanding the expression $U(\phi)$ in a power series in $\phi$ yields
\begin{gather}
U(\phi)=-\frac{1}{2}\left(1-\frac{\mu_1}{M^2-3\theta_1}-\frac{\mu_2}{QM^2-3\theta_2}\right)\phi^2+\nonumber\\
+\frac{1}{6}\left(\frac{3\mu_1(M^2+\theta_1)}{(M^2-3\theta_1)^3}-\frac{3\mu_2(QM^2+\theta_2)}{(QM^2-3\theta_2)^3}-1\right)\phi^3+\\
+\frac{1}{24}\left(\frac{3\mu_1(5M^4+30M^2\theta_1+9\theta_1^2)}{(M^2-3\theta_1)^5}+\frac{3\mu_2(5Q^2M^4+30QM^2\theta_2+9\theta_2^2)}{(QM^2-3\theta_2)^5}-1\right)\phi^4.\nonumber
\end{gather}

In deriving nonlinear equations (21) and (32) (see Sections 3 and 4), we set $\xi=x-Vt$,  where $V$ is the propagation velocity of a linear wave. To find weakly nonlinear solutions, we then passed to the variable $\eta=x-u\tau$, where $u$ is a nonlinear correction to the propagation velocity. Therefore, for the subsequent analysis to be consistent with our previous results, it is necessary to set $M=V+u$, where $u\ll V$. In this case, the expression in the parenthesis in front of $\phi^2$ equals $u/b$ and, in the other terms, it is sufficient to replace $M$ with $V$. As a result, we obtain:
\begin{equation}
\left(\frac{d\phi}{d\eta}\right)^2=\frac{2u}{b}\phi^2-\frac{2A}{3b}\phi^3-\frac{B}{3b}\phi^4.
\end{equation}
The solution to this equation coincides with expression (36).

If the wave properties are determined by the coefficient $A$ and the nonlinear correction $u$ to the propagation velocity is not too small, then the higher order in the expansion of the Mach number in the power series in $u/V$ should be retained in the coefficients of expression (39). In this case, the solution to Eq. (38) can be written as
\begin{equation}
\phi=\frac{3u}{A}\ch^{-2}(\eta/\Lambda)+\frac{9}{2A^2}(4EbV-B/A)u^2\ch^{-2}(\eta/\Lambda)-\frac{9B}{2A^3}u^2\ch^{-2}(\eta/\Lambda)\th^2(\eta/\Lambda),
\end{equation}
where $\Lambda=(2b/u)^{1/2}(1+3\Delta bu/4)$ and
\begin{displaymath}
E=\mu_1\frac{V^2+3\theta_1}{(V^2-3\theta_1)^4}-\mu_2Q\frac{QV^2+3\theta_2}{(QV^2-3\theta_2)^4},\hspace{1 cm}\Delta=\mu_1\frac{V^2+\theta_1}{(V^2-3\theta_1)^3}+\mu_2Q\frac{QV^2+\theta_2}{(V^2-3\theta_2)^3}.
\end{displaymath}
Note that taking into account the next term in the expansion of $M$ is equivalent to constructing a KdV equation that is nonsecular in the second order in $\eps$ and has solutions in the form of the so-called "`dressed"' solitons [13, 14]. From the standpoint of such solutions, the first term in Eq. (41) correspond to the "`core"', whereas the second and the third ones form the "`cloud"'.

For numerical analysis, it is convenient to use the quantity $M'=M/V$ (the Mach number divided by linear propagation velocity (13), which is a root of a biquadratic equation). Far from the critical density of the total negative charge of the dust and ions, there is only one type of solitary wave---either compression or rarefaction one. At the critical density, both types of wave can exist simultaneously and the plot of the quasi-potential $U(\phi)$ intersects the axis @[phi] at the points that are symmetric relative to $\phi=0$. When $M'$ is close to unity, the amplitudes of these waves are small and are well described by formula (33). As  $M'$ ncreases, the amplitudes of both positive and negative waves of the potential increase. Figure 4 illustrates how the shape of the quasi-potential well for the critical values $\alpha=0.202$ and $\beta=0.2$ (all the other parameters are specified in the figure caption) varies with increasing Mach number. As the Mach number increases, the wave amplitude increases and, at a certain value $M'_{cr}$ (for the parameters corresponding to Fig. 4, it is $M'_{cr}\approx 1.129$) , the kinetic energy of an incoming ion in the soliton frame of reference becomes less than its potential energy at the center of the soliton; i.e., the ion is reflected. Figure 5 shows the shape of the quasi-potential well for the case where the dust density (or the negative ion density) is not equal to its critical value. As an example, the figure illustrates the situation in which the negative ion density is constant, whereas the dust density increases and the wave transforms into a rarefaction soliton ($\phi_m<0$). Slight variations in $\beta$  leads to insignificant variations in the amplitudes of compression and rarefaction solitons. For small-amplitude solitons, this is clearly seen from formula (39). If $A^2\ll 6Bu$, then the soliton amplitude is $\phi_m\approx \pm(6u/B)^{1/2}(1\mp A/\sqrt{6Bu})$. In the case at hand, we have $A<0$; therefore, the amplitude of the compression soliton is slightly higher than that of the rarefaction one. It can be seen from the figure that, for waves of arbitrary amplitude, the difference between the amplitudes of the compression and rarefaction solitons becomes considerable. As the dust density increases, the amplitude of a positive-potential soliton becomes higher than $M'^2V^2/2$; i.e., the condition for the existence of such solitons is violated. In this case, the quasi-potential has only one turning point, which corresponds to a rarefaction wave. As the dust grain charge decreases below its critical value, the amplitude of a negative-potential soliton increases and the turning point of the quasi-potential for such solitons also disappears when $\beta$ decreases below a certain value. Note that the range of the values of the dust-density parameter $\beta$ within which both types of solitary waves exist simultaneously is rather wide. Thus, in the case at hand, the parameter $\beta$ can vary in the range $0.159-0.42$. If $\beta$ is lower than $-0.159$ (which corresponds to a positively charged dust), then only positive-potential solitons can propagate in the plasma. In contrast, if $\beta$ is higher than $0.42$, there are only solutions in the form of rarefaction solitons. The table presents the ranges of $\beta$ within which compression and rarefaction solitons can exist simultaneously at different values of $\alpha$.

It should be noted that all the numerical calculations of the wave amplitudes and nonlinear coefficients were performed for the fast ion-acoustic mode [6]. An analysis of the numeric results has shown that the phase velocity of slow ion-acoustic waves is on the order of the ion thermal velocity; therefore, these waves undergo strong Landau damping.

\section{CONCLUSIONS}

Thus, using the approach based on the KdV equation, we have shown that the presence of a negatively charged dust in plasma leads to a decrease in the critical density of negative ions. For a given value of $\alpha$ (the ratio between negative and positive ion densities), the amplitude of a compression soliton $A>0$) in a dusty plasma is larger than that in a bi-ion plasma without dust, whereas the amplitude of a rarefaction soliton ($A<0$) is smaller. In contrast, if the dust is charged positively, then the amplitude of compression (rarefaction) solitons in a dusty plasma is smaller (larger) than that in a plasma without dust.

When the dust is charged positively, there always exists a critical value of the parameter $\alpha$ (the density ratio between the negative and positive ions). For a negatively charged dust, the condition $A=0$ can be satisfied for $\beta$ values below a certain critical value $\beta_{cr}$, at which $\alpha$ becomes zero and above which only rarefaction solitons can propagate in the plasma. In the absence of negative ions, the critical value $\beta_{cr}$ in a cold plasma is equal to $2/3$ [10] and it increases slightly if the thermal effects are taken into account. MKdV solitons in a dusty plasma exist along the corresponding curves in Fig. 2 and only for $\beta<\beta_{cr}$. 

An analysis of the quasi-potential shows that, for waves of arbitrary amplitude, the curve that separates the domains of existence of compression and rarefaction solitons in the plane $\beta-\alpha$ is rather conventional and that there is a broad transition band on both sides of this curve within which both positive- and negative-potential solitons can exist simultaneously. Below this curve (in the domain of existence of compression KdV solitons), the amplitude of rarefaction solitons is larger than that of compression ones. On the other hand, above this curve (in the domain of existence of rarefaction solitons), the amplitude of compression solitons is larger. Above the transition band, only rarefaction solitons exist, whereas below this band, only compression solitons can propagate in the plasma.

\section*{ACKNOWLEDGEMENTS}

I am grateful to Yu.A. Shchekinov for helpful discussions. This work was supported by the Federal Agency for Education (project no. RNP 2.1.1.3483) and the Russian Foundation for Basic Research (project nos. 05-02-17070 and 06-02-16819).

\newpage

\section*{REFERENCES}

[1] Watanabe S.// J. Phys. Soc. Jpn. 1984. V. 53. P. 950.

[2] Tagare S.G.// J. Plasma Phys. 1986. V 36. P. 301.

[3] Verheest F.// J. Plasma Phys. 1988. V. 39. P. 71.

[4] Kalita B.C., Kalita M.K.// Phys. Fluids. 1990. V. B2. P. 674.

[5] El-Labany S., Sheikh A.// Astr. and Space Sci. 1992. V. 197. P. 289.

[6] Mishra M.K., Chhabra R.S.// Phys. Plasmas. 1996. V. 3. P. 4446.

[7] Gill T.S., Kaur H., Saini N.S.// Phys. Plasmas. 2003. V. 10. P. 3927.

[8] McKenzie G.F., Verheest F., Doyle T.B. and Hellberg M.A.// Phys. Plasmas. 2004. V. 11. P. 1772.

[9] McKenzie G.F., Verheest F., Doyle T.B. and Hellberg M.A.// Phys. Plasmas. 2005. V. 12. P. 102305.

[10] Shukla P.K., Mamun A.A.. Introduction to Dusty Plasma Physics. Bristol: Institute of Physics Publishing, 2002.

[11] Popel S.I., Yu M.Y., Tsytovich V.N.// Phys. Plasmas. 1996. V. 3. P. 4313.

[12] Popel S.I., Golub' A.P., Losseva T.V. et al.// Phys. Rev. E. 2003. V. 67. P. 056402.

[13] Kodama Y., Taniuti T.// J. Phys. Soc. Jpn. 1978. V. 45. P. 298.

[14] Tiwari R.S., Mishra K.// Phys. Plasmas. 2006. V. 13. P. 062112.

\newpage
\section*{TABLE}
\begin{flushleft}
\begin{tabular}{|l|l|l|l||l|l|l|l||l|l|l|l|}\hline
$Q$ &$\alpha$&$\beta_{min}$&$\beta_{max}$&$Q$ &$\alpha$&$\beta_{min}$&$\beta_{max}$&$Q$ &$\alpha$&$\beta_{min}$&$\beta_{max}$\\\hline\hline
1 & 0.1 &  0.334 &  0.588& 4& 0.1 & 0.332 & 0.624 &32& 0.1 & 0.100 & 0.639\\
  & 0.2 & -0.159 &  0.420&  & 0.2 & 0.132 & 0.514 &  & 0.2 &-0.069 & 0.536\\
  & 0.3 & -0.902 &  0.251&  & 0.3 &-0.077 & 0.400 &  & 0.3 &-0.240 & 0.436\\
  & 0.4 & -2.204 &  0.078&  & 0.4 &-0.290 & 0.290 &  & 0.4 &-0.408 & 0.335\\
  & 0.5 & -4.113 & -0.115&  & 0.5 &-0.520 & 0.177 &  & 0.5 &-0.578 & 0.232\\\hline
\end{tabular}
\end{flushleft}

Ranges of the parameter $\beta$ within which both compression and rarefaction solitons can exist for $\theta_1=\theta_2=0.01$, $M'=1.05$, and different values of $Q$ and $\alpha$.

\newpage
\section*{FIGURE CAPTIONS}
Fig. 1. Amplitudes $\phi_m$ of (a) compression ($A>0$) and (b) rarefaction solitons ($A<0$) as functions of the parameter $\alpha = n_{i20}/n_{i10}$ for $Q=32, \theta_1=\theta_2=0.05$, $u=0.01$ , and different values of $\beta$.

Fig. 2. Curves corresponding to the condition $A=0$ in the plane ($\beta - \alpha$)  for $\theta_1=\theta_2=0.05$, $u=0.01$, and different mass ratios between the negative and positive ions.

Fig. 3. Amplitude of an MKdV soliton vs. $\beta$ for $\theta_1=\theta_2=0.05$, $u=0.01$, and different values of $Q$.

Fig. 4. Shape of the quasi-potential well at the critical density of negative ions for $\alpha=0.202$, $\beta=0.2$, $Q=1$, $\theta_1=\theta_2=0.01$ , and different values of the Mach number $M$: $1-1.03, 2-1.05, 3-1.1$.

Fig. 5. Shape of the quasi-potential well for $\alpha=0.2$, $Q=1$, $\theta_1=\theta_2=0.01$, $M=1.05$, and different values of $\beta$: $1-0.15, 2-0.25, 3-0.35$.

\newpage

\begin{figure}\center
\includegraphics[height=6.5cm]{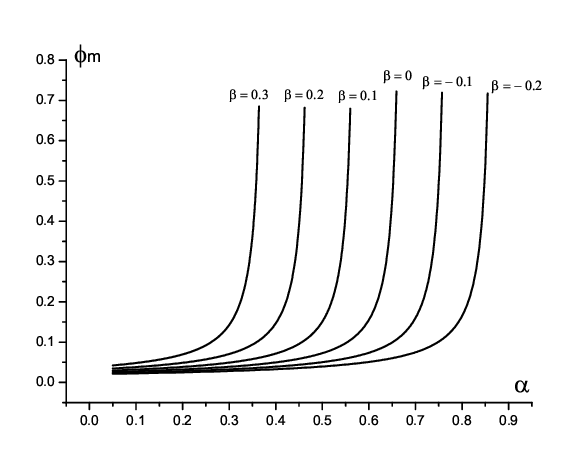}
\includegraphics[height=6.5cm]{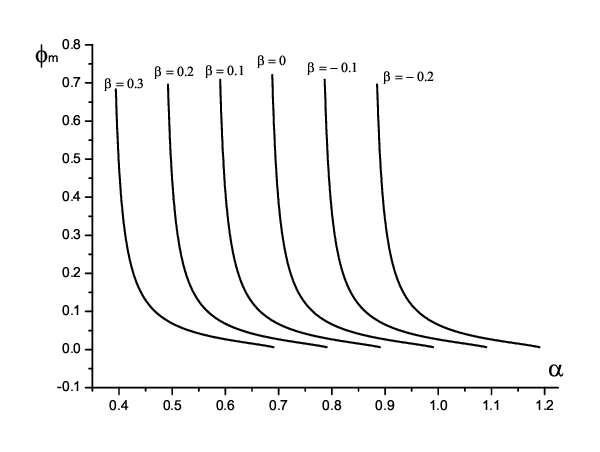}
\\Fig 1.
\end{figure}

\begin{figure}\center
\includegraphics[height=7cm]{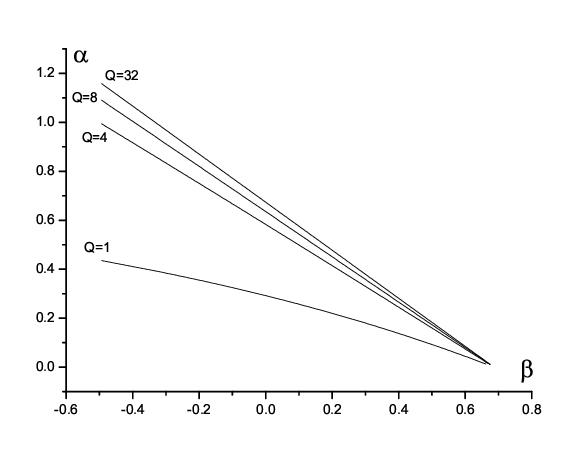}
\\Fig 2.
\end{figure}

\newpage

\begin{figure}\center
\includegraphics[0,0][300,300]{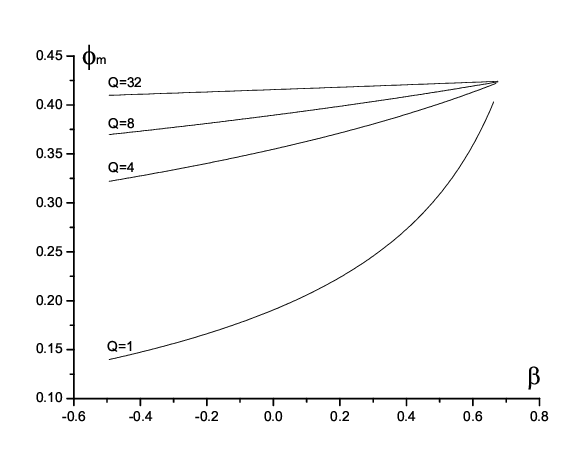}
\\Fig 3.
\end{figure}

\begin{figure}\center
\includegraphics[height=7cm]{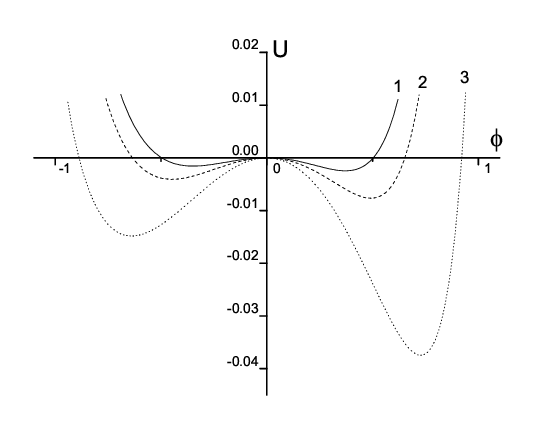}
\\Fig 4.
\end{figure}

\begin{figure}\center
\includegraphics[height=7cm]{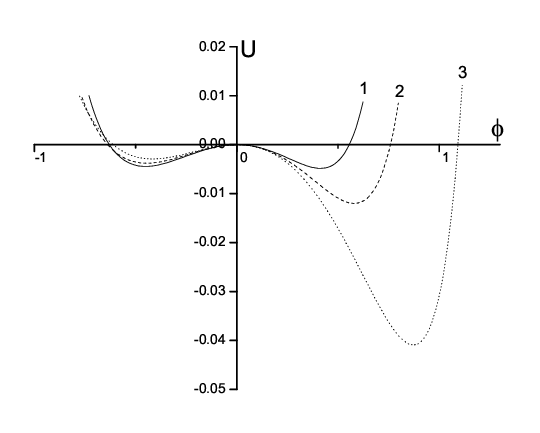}
\\Fig 5.
\end{figure}

\end{document}